\documentstyle[subeqnarray,preprint,pre,aps,eqsecnum,epsf,graphicx]{revtex}

\newcommand{\real}{{\sf I}\kern-.12em{\sf R}}
\newcommand{\comp}{{\sf I}\kern-.50em{\sf C}}
\newcommand{\unity}{{\sf I}\kern-.54em{\sf 1}}
\def\spose#1{\hbox to 0pt{#1\hss}}
\def\ltapprox{\mathrel{\spose{\lower 3pt\hbox{$\mathchar"218$}}
 \raise 2.0pt\hbox{$\mathchar"13C$}}}
\def\gtapprox{\mathrel{\spose{\lower 3pt\hbox{$\mathchar"218$}}
 \raise 2.0pt\hbox{$\mathchar"13E$}}}

\begin{document}

\rightline{IFUP--TH/2005--36}
\rightline{GEF-TH-2005-11}

\vskip 2cm

\centerline{\bf Behaviour of the topological susceptibility in}
\centerline{\bf two colour QCD across the finite density transition}
\vskip 5mm
\centerline{B. All\'es$^a$, M. D'Elia$^{b}$, M. P. Lombardo$^{c}$}
\centerline{\it $^a$INFN, Sezione di Pisa, Pisa, Italy}
\vskip 2mm
\centerline{\it $^b$Dipartimento di Fisica, Universit\`a di Genova and INFN,
Genova, Italy}
\vskip 2mm
\centerline{\it $^c$INFN, Laboratori Nazionali di Frascati, Frascati, Italy}

\begin{abstract}
The behaviour of the topological susceptibility $\chi$ in QCD with two
colours and 8 flavours of quarks is studied at nonzero temperature
on the lattice across the finite
density transition. It is shown that the signal of $\chi$ drops
abruptly at a critical chemical potential $\mu_c$,
much as it happens at the finite
temperature and zero density transition. The Polyakov loop and the
chiral condensate undergo their transitions at the same critical
value $\mu_c$. At a value $\mu_s$ of the chemical potential, called
saturation point, which in our case 
satisfies $\mu_s > \mu_c$, Pauli blocking
supervenes and consequently the theory becomes quenched.
\end{abstract}

\vskip 5mm

\vfill\eject

\section{Introduction}

\vskip 5mm

The topological susceptibility $\chi$ has many applications in QCD.
It is defined as the correlation function of two topological charge density
operators $Q(x)$ at zero momentum where $Q(x)$ is
\begin{equation}
 Q(x)=\frac{g^2}{32 \pi^2}\epsilon_{\zeta\nu\gamma\sigma} F^a_{\zeta\nu}(x)
F^a_{\gamma\sigma}(x)\,,
\label{Qcontinuum}
\end{equation}
and $F^a_{\zeta\nu}(x)$ is the gluon tensor. The quenched value of $\chi$
is responsible for the large mass of the $\eta'$
meson~\cite{witten,veneziano} and allows to understand the realization
of the U$_{\rm A}$(1) symmetry. From phenomenology this quantity must be
$\chi\approx\left(\hbox{180 MeV}\right)^4$. The most recent lattice
results~\cite{lucini,deldebbio,alles1} and calculations based on
SVZ--sum rules~\cite{narison} are in fair agreement with this prediction.
Besides, in the full theory $\chi$ is proportional to the chiral condensate in
the massless limit~\cite{hansen} while
the second moment of the correlation function of two topological
charge density operators furnishes a possible explanation to the so--called
proton spin crisis~\cite{shore}.

The behaviour of the topological susceptibility at finite temperature
and/or matter density is an
important ingredient to understand whether or not the singlet chiral
symmetry undergoes some sort of effective restoration at extreme
conditions. The fate of
this symmetry and of the non--singlet chiral symmetry and the
ordering of the respective phase transitions
open several scenarios~\cite{pis,shuryak,meggiolaro,costa} with relevant
physically measurable consequences.

It is known that $\chi$ undergoes an abrupt
drop at the deconfinement transition at finite temperature in
Yang--Mills theory, both for two~\cite{alles2}
and three~\cite{alles3} colours, and in full QCD for three
colours and 2 or 4 flavours of staggered quarks~\cite{alles4}. These
results have been confirmed by many
collaborations by using cooling~\cite{lucini,teper,ddpv}\footnote{Older
papers claimed that the signal dissapeared smoothly and it began to
fade away well before the deconfinement temperature; however nowadays
this possibility seems excluded.} (see however~\cite{meggiopana})
and by counting fermionic zero modes~\cite{gattringer},
including studies for $N\geq 4$ colours.

In this paper we present an investigation about
the behaviour of $\chi$ at finite temperature and
density. This study has been carried out considering two colour QCD. In this case we do
not have to face the hard sign problem of numerical simulations in three colour QCD at finite
density (although several techniques to overcome these difficulties
have been proposed~\cite{fodor,muim,taylor}).

Several analyses of two colour QCD at finite density are available in the
literature~\cite{Hands99,Kogut01,Kogut00,Aloisio,Muroya,Muroya2,Lombardo3,San,Metlitski}
and one expects that the qualitative aspects of the behaviour of gluonic observables
can be extended to three colour QCD. In particular, for the topological susceptibility, this
expectation is further supported by the fact that instantons~\cite{thooft} are present in any SU($N$)
gauge group theory. Monte Carlo results
bear out this expectation both for the validity of the
Witten--Veneziano mechanism as well as for the
sudden drop of the signal of $\chi$ at the deconfinement
transition~\cite{lucini,alles2,ddpv}.

Model studies suggest that instantons form structures with different features as the
temperature and density change~\cite{rapp,zz}. Beyond the finite density
transition instantons seem to cluster into polymers while at the deconfinement
temperature and zero density they tend to form instanton--antiinstanton pairs.

We have investigated the behaviour of $\chi$ on the lattice
as the density of matter is varied at finite temperature.
The temperature was introduced as usual by compactifying the
temporal direction of the lattice and the density by switching on a
chemical potential $\mu$ for quarks. Fermions have been discretized
by using the staggered formulation~\cite{ks}

The two main results of our numerical study are: {\it i)} the
topological susceptibility drops at a critical $\mu_c$ and {\it ii)}
this $\mu_c$ is the same value of the chemical potential where the
chiral condensate falls and the Polyakov loop rises.
Moreover the introduction of a density of matter brings forth the
following effect.
The density of fermions grows as the value of $\mu$ is increased up to a saturation point,
$\mu_s >\mu_c$, where no more fermions can be inserted into the system.
In fact, due to the Pauli exclusion principle,
there cannot be more than $N_f/4$ fermions per site
for SU(2) gauge group and $N_f$ flavours (which in our case is
equivalent to $N_f/8$ baryons per site). This number follows from
the fact that in the staggered formulation, one Dirac spinor
is reconstructed from the fields at the 16 sites of
every hypercube~\cite{kstern} and that we are considering two colours.
When the system is completely
saturated with fermions, any further increase of the chemical potential will be totally
ineffective: the fermions are completely frozen and the gluons are the only dynamical
degrees of freedom. Then the observables take on their quenched values at $\mu >\mu_s$ and,
in particular, the topological susceptibility grows again until reaching a second plateau.
In fact the corresponding temperature for the quenched theory lies in the confined phase
where $\chi$ is nonzero~\cite{alles3}.
However it is important to notice that the saturation point is an
unphysical effect related to the countable number of sites in the
lattice or, stated otherwise, to the presence of an ultraviolet (UV)
cutoff, and is thus expected to disappear in the continuum limit; an
essential
point is that in our case the saturation point $\mu_s$ turns out to be
larger than the critical chemical potential $\mu_c$ and also well
separated from it, so that the study of the deconfinement phase transition
is not affected by that lattice artifact.

The paper is organized as follows:
in Section~2 the technical details of the numerical simulations are
given; the calculation of $\chi$ and the investigation of its behaviour
across the transition are presented in Section~3 together with the
extraction of the physical units. Section~3 also contains a study of
simultaneity of the chiral, confinement and topological transitions.
Some conclusive comments are left for the last Section.

\section{Monte Carlo simulations}

\vskip 5mm

We have performed numerical simulations in QCD with dynamical
staggered quarks~\cite{ks}. Following Ref.~\cite{Hands99} we have
worked with two colours and 8 flavours of staggered fermions. The action
is~\cite{hasenfratzkarsch}
\begin{eqnarray}\label{action}
 S &=& \frac{1}{2} \sum_x \eta_4(x) \Big\{
                  {\rm e}^{a\mu} \overline{\psi}(x)
                  U_4(x)\psi(x+\widehat{4})
   - {\rm e}^{-a\mu} \overline{\psi}(x+\widehat{4})
                  U_4(x)^\dagger\psi(x) \Big\} \nonumber \\ \nonumber \\
   &&\negthinspace\negthinspace\negthinspace\negthinspace
      + \frac{1}{2} \sum_{i=1}^3\sum_x \eta_i(x) \Big\{
                  \overline{\psi}(x)
                  U_i(x)\psi(x+\widehat{i}\,)
   - \overline{\psi}(x+\widehat{i}\,)
                  U_i(x)^\dagger\psi(x) \Big\} \nonumber \\ \nonumber \\
   && \negthinspace\negthinspace\negthinspace\negthinspace
      + am \sum_x \overline{\psi}(x) \psi(x) + S_W\, ,
\label{actionS}
\end{eqnarray}
where $\eta_\nu(x)$ are the staggered phases and $U_\nu(x)$ are the
gauge links starting at site $x$ and directed towards $\widehat{\nu}$.
$S_W$ is the usual Wilson action for gauge fields~\cite{wilson}. Fermion
fields have been multiplied by a suitable power of
the lattice spacing $a$ in order to render them dimensionless.

 For further details on the algorithm and the action we refer the
reader to Ref.~\cite{Hands99}. We remind the reader that the
chiral symmetries of the lattice action are not equal to the ones of the
corresponding continuum theory.

We have numerically simulated the model on a
$14^3\times 6$ lattice at inverse bare gauge coupling
$\beta\equiv4/g^2=1.5$ and bare quark mass $am=0.07$ by using
the Hybrid Molecular Dynamics (HMD) algorithm~\cite{duanekogut}.
The presence of a chemical potential prevents us from
separating odd and even sites~\cite{hmc} and as a consequence
we worked with 8 flavours. We have not used the
square root trick to reduce the number of quark species.

We have run the Molecular Dynamics algorithm with step size
$\delta\tau=0.02$~\cite{Hands99} and 50
decorrelation HMD steps have been performed between
two successive measurements. This separation was chosen so to have
configurations with well--decorrelated topological
properties~\cite{boyd,lippert,aokiQ}. For a few values
of $\mu$, 100 HMD steps were used.

Although we have used a non--exact algorithm, possible systematic
errors due to the finite step size seem to be under control. This
statement is
corroborated by two facts: {\it i)} runs corresponding to the same
input parameters but with finer HMD step size ($\delta\tau=0.005$)
have been carried out and results remain stable;
{\it ii)} at large
chemical potential, where the theory becomes effectively quenched,
our results are in complete agreement with similar measurements obtained
from separate runs performed by simulating the pure gauge action,
(see Section~3.c).

We have calculated the topological susceptibility from the
fluctuations of the topological charge, estimated by using
the 1--smeared operator (see next Section). We also measured~the

\begin{figure}
\centerline{\includegraphics[width=\columnwidth]{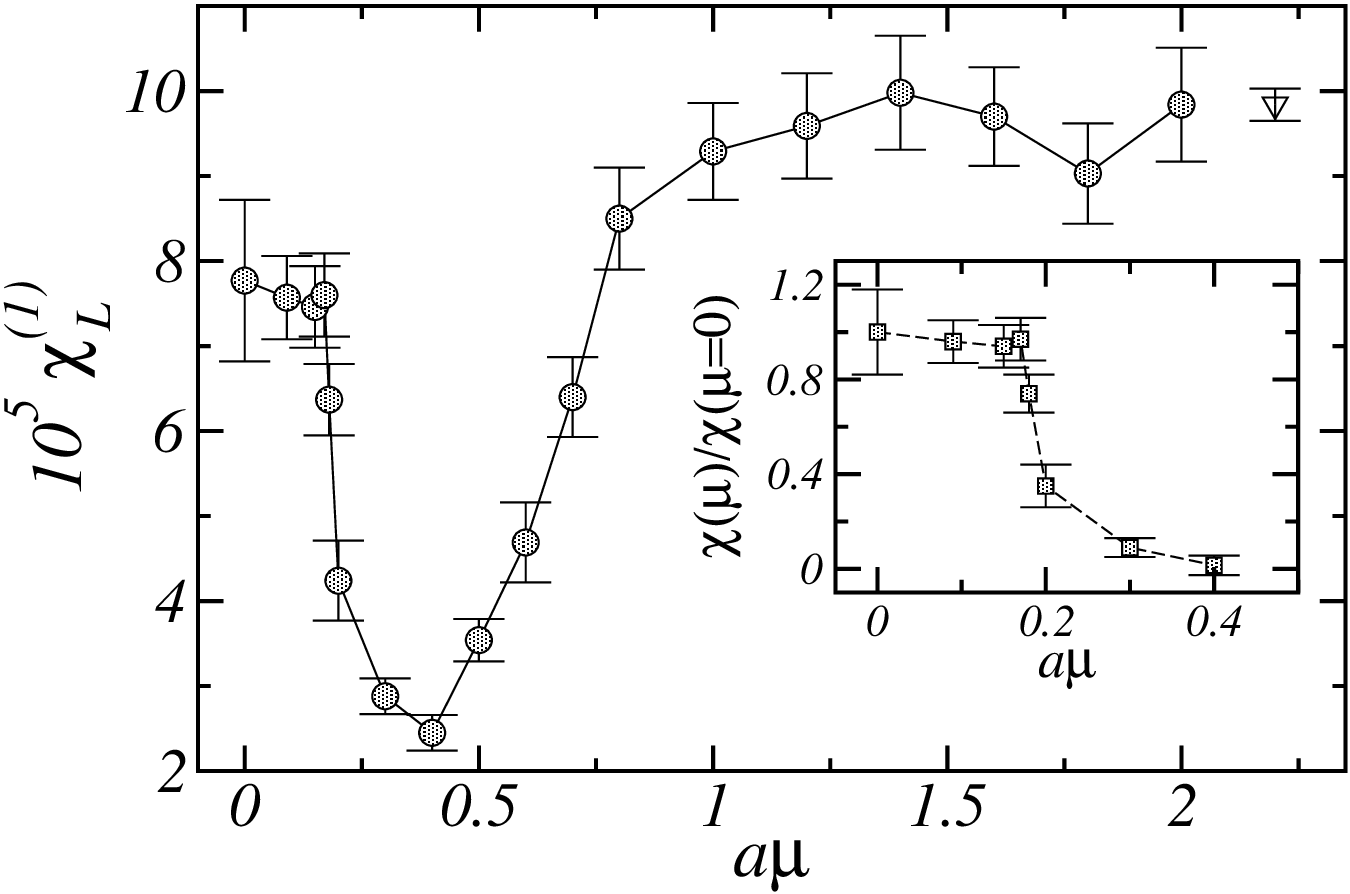}}
\vspace*{12pt}
\caption{$\chi^{(1)}_L$ (circles) as a function of the adimensional
chemical potential $a\mu$. The last point (empty down triangle) is the
value of $\chi_L^{(1)}\Big\vert_{\rm quenched}$ at the same $\beta$. In the
inset the ratio $\chi(\mu)/\chi(\mu=0)$ (squares) is shown in the interval
of $a\mu$ where no apparent effects of Pauli blocking are detectable.
Lines are drawn to guide the eye.}
\end{figure}

\noindent Polyakov loop,
the average plaquette, the baryon density and the chiral condensate.
Depending on the value of $a\mu$, the statistics for these
observables was 300--500 uncorrelated measurements.

Early exploratory runs indicate that on a $14^3\times 6$
lattice at $\beta=1.5$, $am=0.07$ and zero chemical potential,
the system is in the normal hadronic phase.
According to the suggested phase diagram for
two colour QCD in the temperature--density
plane (Fig.~1 of Ref.~\cite{Kogut02}),
our simulations will cross a transition line
which separates the hadronic phase from another state.
Depending on the temperature, this state can be either
the quark--gluon plasma or
the superfluid state and the
respective transition can change its nature correspondingly.
In

\begin{figure}
\centerline{\includegraphics[width=\columnwidth]{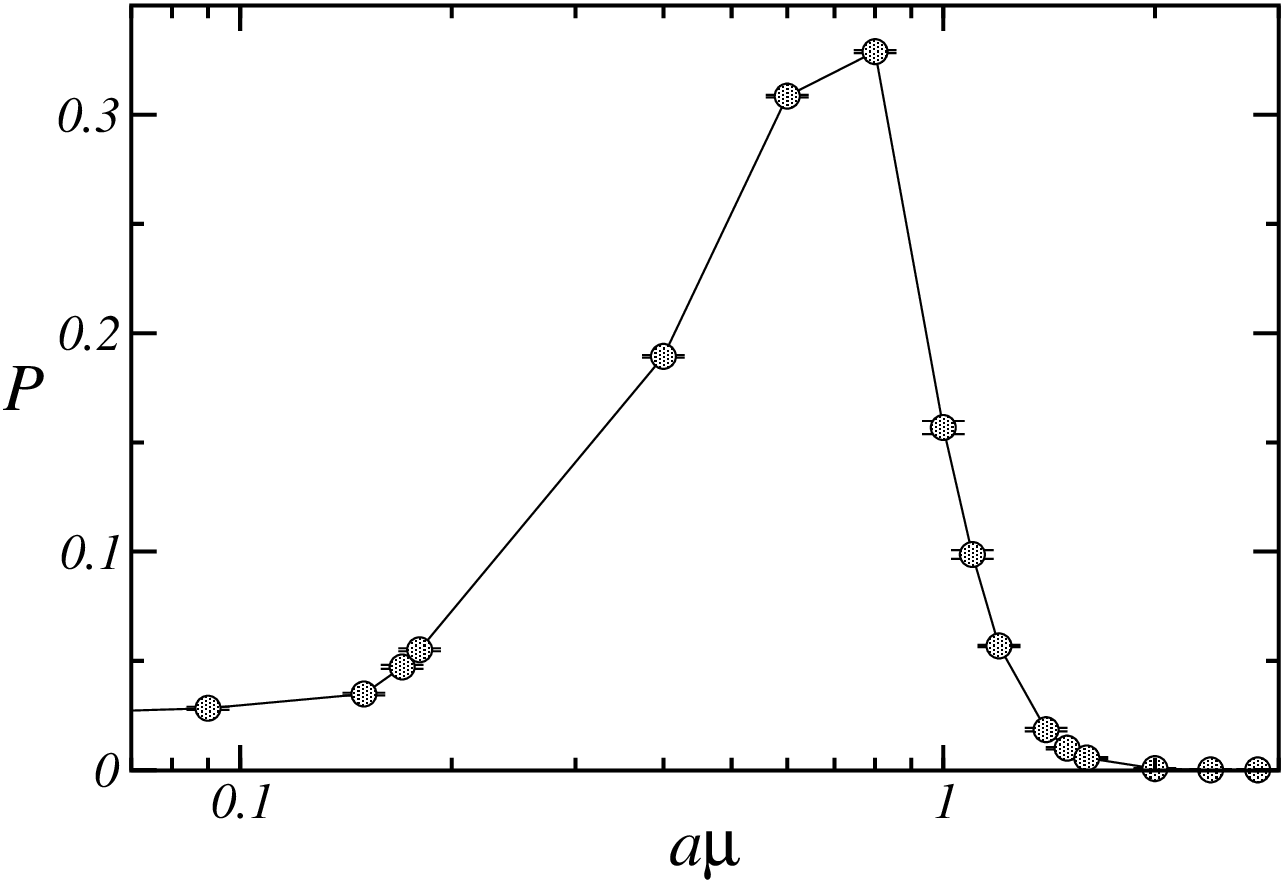}}
\vspace*{12pt}
\caption{Polyakov loop $P$ as a function of $a\mu$.
The logarithmic scale allows to disentangle the data
obtained in the vicinity of the transition point.
Points are joined by a line to guide the eye.}
\end{figure}

\noindent the present work we have not examined the order of
this transition but we have shown that the transitions for the
Polyakov loop, chiral condensate and topological susceptibility are
concomitant. From this fact and the calculated value of the ratio
$T/T_c$ (see later) we infer that possibly our system becomes a
deconfined plasma of quarks and gluons for $\mu >\mu_c$ ($T_c$
is defined as the temperature where the chiral vector symmetry is restored
at zero chemical potential). Moreover the fermionic saturation
is consistently observed at roughly the same value $\mu_s$ for all observables.

\section{Measurement of $\chi$ and its behaviour across the transition}

\vskip 5mm

\subsection{The method}

The first task to calculate the topological susceptibility on the
lattice is the introduction of a regularized topological charge
density operator $Q_L(x)$. We require that this operator satisfies
the continuum limit (in the operator sense)
$Q_L(x){\buildrel a\rightarrow 0 \over \longrightarrow}a^4 Q(x)$
where $a$ is the lattice
spacing. By calling $Q_L\equiv\sum_x Q_L(x)$ the
lattice topological susceptibility is defined as
\begin{equation}
\chi_L\equiv\frac{\Big\langle\left(Q_L\right)^2\Big\rangle}{V}\,,
\label{chiL1}
\end{equation}
where $V$ is the spacetime volume of the lattice. This quantity is
not yet equal to the physical susceptibility $\chi$.
They are related through the general expression~\cite{camponpb329}
\begin{equation}
\chi_L=Z^2\,a^4\,\chi + M\,,
\label{chiL1bis}
\end{equation}
where $Z$ and $M$ are renormalization constants usually called
multiplicative and additive respectively. They depend on the
lagrangian (specifically on the quark masses and the gauge coupling)
and on the operator used on the lattice to regularize
the topological charge.
In particular had we used an operator $Q_L(x)$
suggested by a fermion theory obeying the Ginsparg--Wilson
condition~\cite{gw,neuberger} then $Z$ would have become
equal to 1 and $M$ equal to zero~\cite{giusti}.
We did not make use of such operators because their calculation
on the lattice is very demanding in computer time.

In order to obtain $\chi$ from Eq.~(\ref{chiL1bis}) we have used
the Pisa method (also called ``field--theoretical method''). In this
method one first calculates $Z$ and $M$ and then inserts their values in
Eq.~(\ref{chiL1bis}), to extract the result for $a^4\chi$.

Let us explain the origin of $Z$ and $M$ in Eq.~(\ref{chiL1bis}).
Let $G^{(n,l)}(p_1,p_2,\cdots;q_1,q_2\cdots)$ be a bare Green's function with
$n$ elementary fields (carrying momenta $p_i$) and $l$ insertions
of the topological charge density operator (carrying momenta $q_i$) to
be computed in the quenched theory.
This function can be calculated either in the continuum by using some
regularization method or on the lattice. In both cases the inclusion
of renormalization constants for the external fields and gauge
coupling is enough to make the Green's function finite (because the
calculation is performed in the quenched theory where $Q(x)$ is a
renormalization group invariant operator). Let us call
$G^{(n,l)}_{R,{\rm lattice}}$ and $G^{(n,l)}_{R,{\rm continuum}}$ the
two renormalized functions thus obtained.
In general, for an arbitrary renormalization scheme, the two functions
$G^{(n,l)}_{R,{\rm lattice}}$ and $G^{(n,l)}_{R,{\rm continuum}}$
are not necessarily equal and
they match only after the inclusion of a multiplicative
finite renormalization constant $Z$~\cite{campoharis,allesvicari}
for each of the $l$ insertions of topological charge operator.
In a formal writing: $Q_L(x)=Z a^4 Q(x)$. When $l>1$ there may be further
divergences originated by contact terms, see later.

In the theory with fermions the topological charge mixes with
other operators related to the axial anomaly~\cite{espriu}. This mixing
induces a correction to the above described multiplicative renormalization.
Such a correction is however rather small~\cite{alles5} and we neglect it,
consistently with the large statistical and systematic errors of our
numerical simulation.

The r.h.s. of Eq.~(\ref{chiL1}) contains the product of two
operators at the same spacetime
point. In Field Theory such products
can be divergent~\cite{zimmerman}. On the other hand the correlator
$\langle Q(x) Q(0)\rangle$ is negative for $x\not= 0$
as shown in~\cite{seiler,osterwalder,menotti}.
Also on the lattice this correlator is
negative~\cite{kirchner,horvath,ilgenfritz}.
Then part of the contact divergences
must add up to the total topological susceptibility in order to make
it positive. The rest of the contact divergences, if any, must
be subtracted.
This subtraction is $M$. In order to calculate it, we can follow the strategy
introduced in~\cite{gunduc2284,alles3}: $M$ is the value
of $\chi_L$ in the sector of zero total topological charge,
$M\equiv\chi_L\Big\vert_{Q=0}$.
This prescription guarantees the physical requirement that
$\chi$ vanishes in such a sector.
$M$ can be also calculated in nonzero topological
charge sectors~\cite{massimo}
yielding concordant results.
Although $M$ consists of UV divergences,
it turns out to be a finite quantity because we are working on the
regularized theory (in the practical calculation
the cutoff $a$ is never sent to zero).

We have calculated the topological susceptibility for several values
of the chemical potential at a fixed inverse gauge coupling $\beta$.
This entails that the additive and multiplicative renormalization constants
must be the same for all values of $\mu$. Since we will present the~res-


\vfill\eject


\centerline{{\bf Table 1}}
\centerline{Measurements of several operators for varying $\mu$.}
\vskip 2mm
{\centerline{
\vbox{\offinterlineskip
\halign{\strut
\vrule \hfil\, $#$ \hfil \,&
\vrule \hfil\, $#$ \hfil \, &
\vrule \hfil\, $#$ \hfil \, &
\vrule \hfil\, $#$ \hfil \, &
\vrule \hfil\, $#$ \hfil \, &
\vrule \hfil\, $#$ \hfil \, &
\vrule \hfil\, $#$ \hfil \, &
\vrule \hfil\, $#$ \hfil \, \vrule \cr
\noalign{\hrule}
a\mu &
10^5\,\chi^{(1)}_L &
10^5\,M &
a^3\rho_B &
\langle {1 \over 2}{\rm Tr}\Box\rangle &
\langle \overline{\psi}\psi\rangle_{\rm full} &
\langle \overline{\psi}\psi\rangle_{\rm quenched} &
P \cr
\noalign{\hrule}
0.0 &
 7.77(95)&
 2.54(19)&
 0.001(1) &
 0.4621(2)&
0.592(2) &
0.904(3) &
0.023(1) \cr
\noalign{\hrule}
0.09 &
7.57(49) &
- &
- &
0.4627(3) &
0.587(2) &
- &
0.0283(7) \cr
\noalign{\hrule}
0.15 &
7.46(48) &
- &
- &
0.4631(2) &
0.581(2) &
- &
0.0348(6) \cr
\noalign{\hrule}
0.17 &
7.60(49) &
 -&
0.005(1) &
0.4647(4) &
0.568(2) &
- &
0.0472(9) \cr
\noalign{\hrule}
0.18 &
6.37(42) &
 -&
0.006(2) &
0.4665(2) &
0.552(2) &
- &
0.0551(7) \cr
\noalign{\hrule}
0.2 &
 4.24(47)&
 2.29(19)&
- &
- &
- &
- &
- \cr
\noalign{\hrule}
0.3 &
 2.88(21)&
 -&
 -&
 -&
 -&
- &
- \cr
\noalign{\hrule}
0.4 &
 2.45(21)&
 2.16(13)&
0.060(2) &
0.4912(2) &
 0.280(1)&
0.493(3) &
0.1894(6) \cr
\noalign{\hrule}
0.5 &
 3.54(25)&
 -&
 -&
- &
- &
- &
- \cr
\noalign{\hrule}
0.6 &
 4.69(47)&
 2.73(19)&
 0.230(2)&
0.4764(2) &
0.105(1) &
- &
0.3085(6) \cr
\noalign{\hrule}
0.7 &
 6.40(47)&
- &
- &
- &
- &
- &
- \cr
\noalign{\hrule}
0.8 &
8.50(60) &
- &
0.554(2)&
0.415(2) &
0.062(1) &
0.059(9) &
0.3289(7) \cr
\noalign{\hrule}
1.0 &
 9.29(57)&
- &
0.882(2) &
0.3742(5) &
0.015(3) &
0.005(1) &
0.157(3) \cr
\noalign{\hrule}
1.1 &
- &
- &
0.958(1) &
0.3668(3) &
0.003(1) &
0.003(1) &
0.099(2) \cr
\noalign{\hrule}
1.2 &
 9.59(62)&
- &
0.987(1) &
0.3638(1) &
 0.0014(3) &
 3(5)\,10^{-4} &
 0.057(1) \cr
\noalign{\hrule}
1.4 &
 9.98(67)&
- &
0.998(1) &
0.3626(4) &
0.0012(4) &
3(3)\,10^{-4} &
0.019(1) \cr
\noalign{\hrule}
1.5 &
- &
- &
0.999(1) &
0.3626(1) &
1(2)\,10^{-4} &
- &
 0.010(1) \cr
\noalign{\hrule}
1.6 &
9.70(58) &
- &
1.000(1) &
0.3625(1) &
-2(2)\,10^{-4} &
-3(2)\,10^{-4} &
 0.006(1) \cr
\noalign{\hrule}
1.8 &
9.03(59) &
- &
- &
- &
- &
- &
- \cr
\noalign{\hrule}
2.0 &
9.83(67) &
- &
1.000(1) &
0.3625(1) &
 3(5)\,10^{-5} &
- &
7(2)\,10^{-4} \cr
\noalign{\hrule}
2.4 &
 -&
- &
0.999(1) &
0.3624(1) &
 6(6)\,10^{-5} &
- &
1(4)\,10^{-4} \cr
\noalign{\hrule}
2.8 &
 -&
- &
1.000(1) &
0.3623(1) &
 3(4)\,10^{-5} &
- &
1(4)\,10^{-4} \cr
\noalign{\hrule}
3.2 &
- &
- &
1.000(1) &
0.3626(1) &
 2(3)\,10^{-5} &
 -1(1)\,10^{-4} &
3(4)\,10^{-4} \cr
\noalign{\hrule}
}}
}}

\vfill\eject


\begin{figure}
\centerline{\includegraphics[width=\columnwidth]{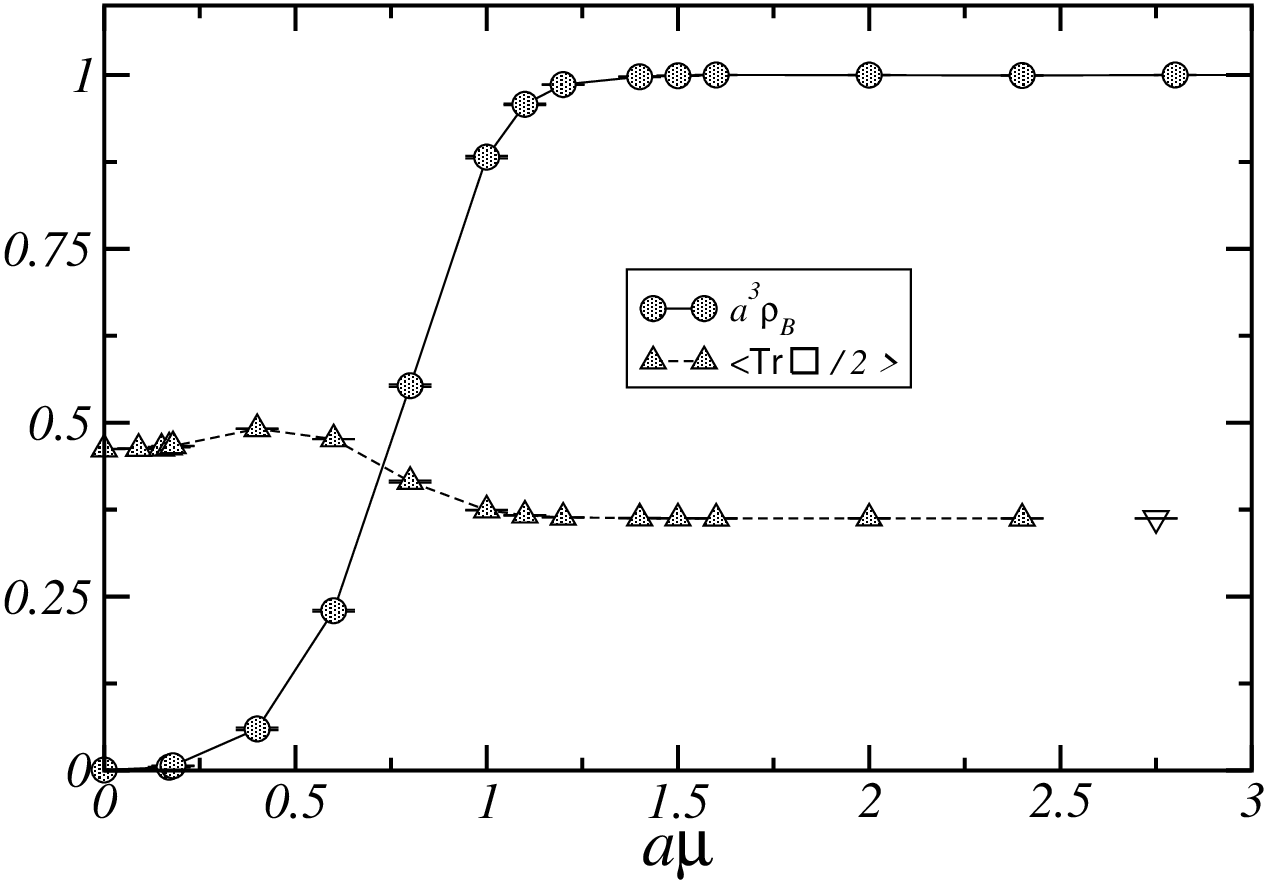}}
\vspace*{12pt}
\caption{Baryonic density $a^3\rho_B$ (circles) and plaquette 
$\langle{1 \over 2}{\rm Tr}\Box\rangle$ (up triangles) as a function of
$a\mu$. The last point (empty down triangle) is the value of the
plaquette for the quenched theory at the same $\beta$ value. Lines
are drawn to guide the eye.}
\end{figure}

\noindent ults for the ratio $\chi(\mu)/\chi(\mu=0)$ as a function of $\mu$,
the multiplicative $Z$ needed not be calculated
(the notation $\chi(\mu)$ indicates the physical topological susceptibility
at a value $\mu$ of the chemical potential). Instead $M$ was calculated
by use of the heating method~\cite{vicdd,gunduc2284}. This method consists
in the following: we started with the trivial
configuration (all links equal to unity) which clearly belongs to the
zero topological sector. Then we applied 5000
steps of HMD updating and measured $\chi_L$ every 100 steps. 
This set of 50 measurements is called ``trajectory''.
After each measurement we applied also 8 cooling
steps~\cite{camponpb329,tepcool,cosmai}
and determined $Q_L$ and the total energy of the configuration
to verify that the topological~sector was not changed (other methods
to check the background topological sector, like the counting of
fermionic zero modes~\cite{pica}, yield the
same results for $M$ and $Z$). We repeated
the whole procedure to obtain a number of trajectories. For 
each trajectory we always discarded
the first few measurements because the configuration was not
yet thermalized\footnote{In this context thermalization is defined
with respect to short range fluctuations within the zero topological
charge sector.}. Averaging over the thermalized steps
(as long as the corresponding cooled configuration showed the correct
topological charge sector, $Q=0$ within a deviation $\delta$)
yielded $\chi_L\Big\vert_{Q=0}$. We estimated
the systematic error that stemmed from the choice 
of $\delta$ as in Ref.~\cite{alles3}.

When the cooling (or other) test reveals that a configuration has
moved from the zero topological charge sector to another sector,
it is discarded. Actually this event seldom happens because the
topological modes are effectively decoupled from the UV modes so that
after starting from a classical configuration of any fixed topological
content, it is difficult to alter the background topological sector
by applying some updating (heating) steps at the corresponding value
of $\beta$ to thermalize the UV fluctuations. In fact this
decoupling of the two types of modes is the gist of the heating
method\cite{vicdd,gunduc2284}.

\subsection{The lattice operator}

A valid lattice regularization of the topological
charge density operator, Eq.~(\ref{Qcontinuum}), is given by~\cite{viejo}
\begin{equation}
 Q_L(x) = -\frac{1}{2^9 \pi^2}\sum_{\zeta\nu\gamma\sigma=\pm 1}^{\pm 4}
 \widetilde{\epsilon}_{\zeta\nu\gamma\sigma} {\rm Tr}\left\{\Pi_{\zeta\nu}(x)
 \Pi_{\gamma\sigma}(x)\right\}\,,
\label{Qlattice}
\end{equation}
where $\Pi_{\zeta\nu}(x)$ is the plaquette in the $\zeta-\nu$ plane with
the four corners at $x$, $x+\widehat{\zeta}$,
$x+\widehat{\zeta}+\widehat{\nu}$, 
$x+\widehat{\nu}$ (counter--clockwise path). As indicated
in the sum, indices $\zeta$, $\nu$, $\gamma$ and
$\sigma$ can point towards either positive or negative directions.
Links pointing to negative directions mean
$U_{-\nu}(x)\equiv U_\nu^\dagger(x-\widehat{\nu})$.
The generalized completely
antisymmetric tensor is defined by $\widetilde{\epsilon}_{1234}=1$ and 
$\widetilde{\epsilon}_{(-\zeta)\nu\gamma\sigma}=-\widetilde{\epsilon}_{\zeta\nu\gamma\sigma}$.

\begin{figure}
\centerline{\includegraphics[width=\columnwidth]{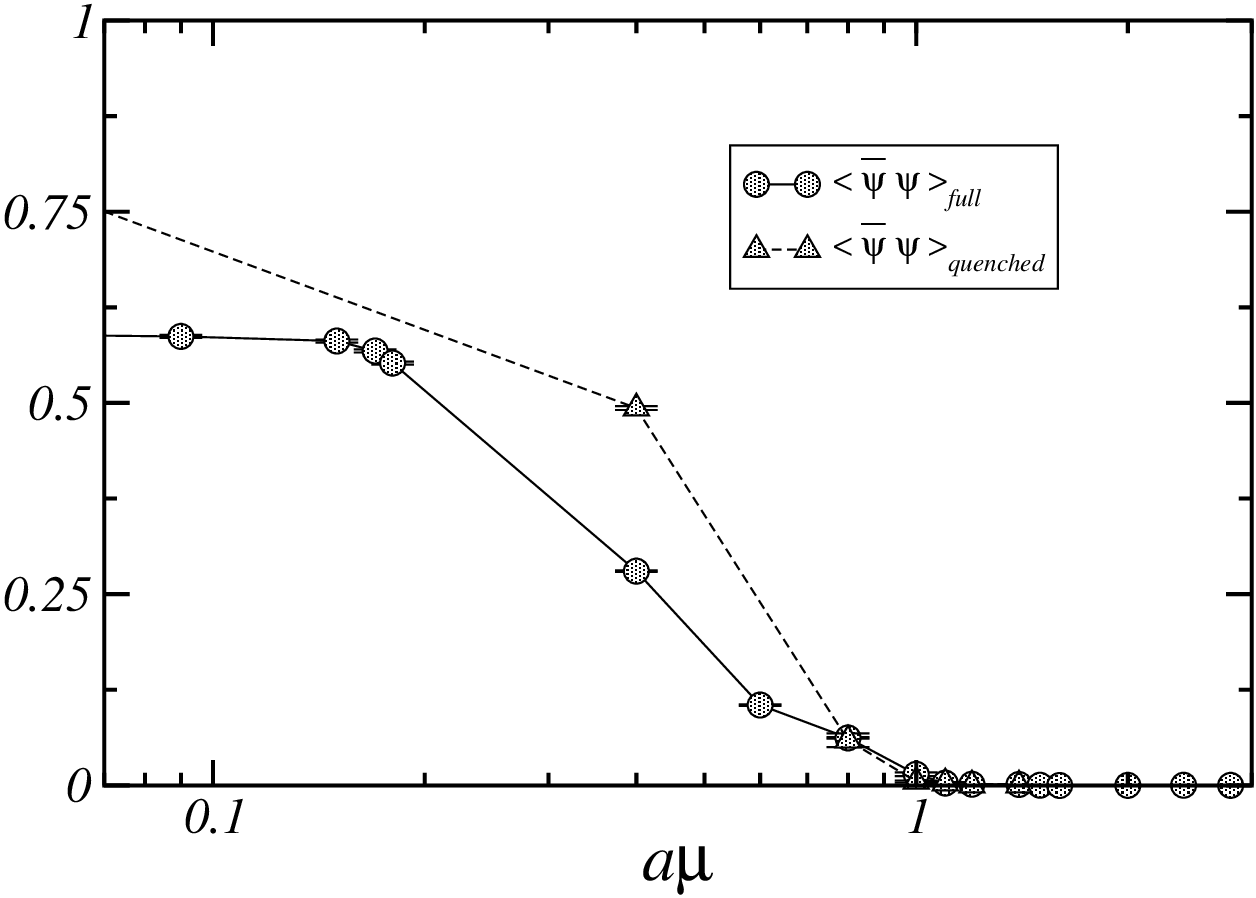}}
\vspace*{12pt}
\caption{Chiral condensate for the full theory (circles) and
the quenched theory (triangles) as a function of $a\mu$.
The logarithmic scale allows to disentangle the data
obtained in the vicinity of the transition point.
Lines are drawn to guide the eye.}
\end{figure}

The lattice operator used in our simulations
was the 1--smeared $Q_L^{(1)}(x)$ which is constructed from
Eq.~(\ref{Qlattice}) after substituting all link matrices
by 1--smeared links~\cite{christou}. If $U_\nu(x)$ is the link
starting at site $x$ and pointing towards the $\widehat\nu$ direction,
the 1--smeared link $U_\nu^{(1)}(x)$ is defined by the following linear combination
of the seven shortest paths connecting the points $x$ and $x+\widehat\nu$
\begin{eqnarray}
 {\overline{U}}_\nu(x) &\equiv& (1-c) U_\nu(x) + \frac{c}{6}
 \sum_{{\scriptstyle \alpha = \pm 1} \atop { \scriptstyle 
 |\alpha| \not= \nu}}^{\pm 4} 
 U_\alpha(x) U_\nu(x+{\widehat\alpha}) U_\alpha^\dagger(x+{\widehat\nu})\,,
 \nonumber \\
 U_\nu^{(1)}(x) &\equiv& \frac{{\overline{U}}_\nu(x)}
 {\left({1\over 2}{\rm Tr} {\overline{U}}_\nu^\dagger(x)
 {\overline{U}}_\nu(x)\right)^{1/2}}\,.
\label{Usmear}
\end{eqnarray}
This combination is then projected back onto SU(2) as
indicated by the second line in Eq.~(\ref{Usmear}).
By using smeared operators less noisy results are obtained.
The parameter $c$ can be tuned in order to optimize this improvement.
We chose $c=0.85$ as in~\cite{alles2}. The na\"\i ve
continuum limit of $Q_L^{(1)}(x)$ equals the continuum topological
charge density, Eq.~(\ref{Qcontinuum}).

{}Following the definition~(\ref{chiL1}) we called
$Q_L^{(1)}\equiv\sum_x Q_L^{(1)}(x)$ the total 1--smeared lattice topological charge and
\begin{equation}
\chi_L^{(1)}\equiv\frac{\Big\langle\left(Q_L^{(1)}\right)^2\Big\rangle}{V}
\label{chiL1111}
\end{equation}
the 1--smeared lattice topological
susceptibility. $\chi_L^{(1)}$
and $a^4\chi$ are related by Eq.~(\ref{chiL1bis}).
We have calculated $M$ for four
values of $\mu$ as shown in Table~1. Averaging over all four results
leads to $M(\beta=1.5,am=0.07)=2.37(8) \times 10^{-5}$. This is
the value of $M$ that we have used throughout the paper.

\subsection{The Monte Carlo simulation}

In Fig.~1 the results for $\chi^{(1)}_L$ are shown as a function of
$a\mu$. Data are given in Table~1. The sudden drop at
$a\mu_c=0.175(5)$ is apparent. In the
inset of Fig.~1 we show the scaling ratio $\chi(\mu)/\chi(\mu=0)$
for $a\mu\leq 0.4$. Accordingly with Eq.~(\ref{chiL1bis})
this ratio is calculated as

\noindent \begin{equation}
{\chi(\mu)\over\chi(\mu=0)}=
\frac{\chi_L^{(1)}(\mu)-M}{\chi_L^{(1)}(\mu=0)-M}\,.
\end{equation}
The notation $\chi_L^{(1)}(\mu)$ stands for the lattice topological
susceptibility, Eq.~(\ref{chiL1111}), calculated at chemical potential $\mu$.
This inset is one of the two main results of our paper: it displays
the drop of the physical topological susceptibility. Notice that the
results in this plot are free from $Z$ and $a^4$ factors.

In Fig.~2 the Polyakov loop is shown in a similar range of values of
chemical potentials at
the same $\beta$ and quark mass. All data are listed in Table~1.
The rise of the Polyakov
loop displayed by this Figure happens at a value of
$a\mu$ which is compatible within errors with the value of
$a\mu_c$ at which $\chi(\mu)$ drops.

For values of the chemical potential larger than $a\mu=0.5$, the topological
susceptibility begins to increase again and then flattens at $a\mu\gtapprox 1.2$, as
shown in Fig.~1. Such a value corresponds to the saturation
point, $\mu_s$. This new plateau is not the indication of a new physical phase.
It is due to the Pauli blocking, a saturation
of the fermion number throughout the lattice.
The value for the chemical potential is so high that the baryon density
$\rho_B$ reaches the maximum allowed value, 1 baryon per site.
After this point the fermions are frozen and the theory is entirely
analogous to the quenched theory. To say it plainly, simulating the
theory at those high values of $\mu$ is nothing but a clumsy way
to reproduce the SU(2) pure gauge theory.

The above qualitative argument is validated by Monte Carlo
simulations of the pure gauge theory that compare very well
with the full theory simulated at $\mu \gtapprox \mu_s$. In
particular the empty down triangle in Fig.~1 is the value of
$\chi_L^{(1)}$ calculated in a separate simulation by using
the SU(2) pure gauge Wilson action, its value being 
$\chi^{(1)}_L\Big\vert_{\rm quenched}=9.84(19)\,10^{-5}$.
The agreement between this number and the last data 
for $a\mu \gtapprox 1.2$ is manifest. Also the signal for the
Polyakov loop in Fig.~2 disappears at $a\mu \gtapprox 1.2$.
This is the expected behaviour for this operator in the quenched
theory when, as in our case, the temperature is below deconfinement,
$T < T_c$.

The rise of the signal of $\chi^{(1)}_L$ for $\mu\ltapprox\mu_s$
is slow in contrast with its sudden drop at $\mu_c$. This effect
is due to the partial blocking of fermions which increases with
$\mu$ until it reaches the total saturation at $\mu_s$.

Notice that the saturation of the fermion degrees of freedom
makes the theory to behave as the pure gauge theory in all respects.
In particular not only $\chi^{(1)}_L$ assumes the quenched value,
but also the renormalization constants and the beta function do (i.e.:
the full theory becomes quenched, also perturbatively),
in such a way that the physical susceptibility $\chi$ turns into that
of the pure gauge theory too.

The saturation of fermions and subsequent forced quenching of the
theory is due to the fact that on the lattice there are a finite
number of physically accessible sites. Hence it is an unphysical
lattice effect with no counterpart in the continuum. It is interesting
to notice that the effective quenching at large values of $\mu > \mu_s$ 
corresponds to the fermionic determinant becoming a 
constant independent of the gauge field background; this can be easily 
understood looking at the fermionic part in Eq.~(\ref{actionS}): in
the large $\mu$ limit the forward temporal part of the Dirac matrix
is exponentially larger than the other terms, and the determinant of a
matrix containing only the forward temporal part is indeed independent
of the gauge field background, as can be easily verified.

In Fig.~3 we show the data for the
baryonic density $a^3 \rho_B$
and the plaquette $\langle{1 \over 2}{\rm Tr}\Box\rangle$ versus the
chemical potential. Both data sets can be found in Table~1.
At the saturation point, $a\mu_s\approx 1.2\,$,
the density attains the maximum value $a^3\rho_B=1$
and the plaquette becomes that of the quenched theory.
This value of $a\mu_s$ is roughly the same at which
the topological susceptibility begins the second plateau in Fig.~1
and the Polyakov loop becomes zero in Fig.~2.
The quenched value of the plaquette at $\beta=1.5$
has been calculated in a separate simulation with the
pure gauge Wilson action and it is shown as
an empty triangle in Fig.~3. It amounts
to $\langle{1 \over 2}{\rm Tr}\Box\rangle_{\rm quenched}=0.362432(23)$
which is in perfect agreement with the (Pauli blocked) values
of the average plaquette in the full theory for $\mu>\mu_s\,$, (see Table~1).

All observables agree with their corresponding quenched values as soon as
the chemical potential is high
enough to induce fermionic saturation.
We have also calculated the chiral condensate as a function of
$a\mu$. The results are shown in Table~1 and Fig.~4 for both the
full and quenched theories. 
In the two simulations, we have measured the same operator, i.e. the
trace of the inverse of the ($\mu$--dependent) quark matrix. In one case
it was calculated by using configurations obtained from the full theory and in the
other case by using configurations corresponding to the pure gauge theory.
In the full theory the condensate drops at the same $a\mu_c$ where
the Polyakov loop rises and the topological susceptibility falls abruptly.
Thus the three operators undergo their transitions at the same density.
We are aware that neither the Polyakov loop nor the topological susceptibility
are order parameters for confinement and the singlet flavour
U$_{\rm A}$(1) symmetry respectively. In particular the drop of the
signal of the topological susceptibility does not necessarily mean
a restoration of the U$_{\rm A}$(1) symmetry although it must have
physically measurable consequences in observables directly related
to~$\chi$. This set of consequences is commonly termed
as ``effective restoration'' of
the U$_{\rm A}$(1) symmetry~\cite{pis,meggiolaro}.

 For values of the chemical potential larger than the
saturation point $\mu_s$ the chiral
condensate in the full theory stays vanishingly small.
This behaviour coincides with that of the chiral condensate
calculated in the quenched theory as shown in Fig.~4.
Notice that in the present case, the operator depends
on the chemical potential and this is why we show the data for
the quenched model as a function of $a\mu$. However as soon as
the chemical potential reaches the saturation value
the two sets of data become coincident.
This is another indication that the full theory at $\mu>\mu_s$
becomes entirely quenched. In the quenched case a straightforward
algebra yields the leading dependence for large $\mu$
\begin{equation}
\langle\overline{\psi}
\psi\rangle_{\rm quenched}\propto{\rm e}^{-N_t a \mu}\, ,
\end{equation}
where $N_t$ is the lattice temporal extent,
$N_t = 6$ in our case.

\subsection{The numerical results}

We have also given approximate physical units to the lattice spacing $a$ and
therefore also to the critical value of the chemical potential $\mu_c$.
In a separate Monte Carlo simulation we have calculated the
chiral condensate and the Polyakov loop as a function of the temperature
at zero chemical potential. To this end we have kept fixed the temporal
lattice size and have varied the gauge coupling $\beta$. In Fig.~5 and Table~2
we show the results of the simulation. From the position of the flex
point in any of the two curves of this Figure we can extract the
critical beta, $\beta_c=1.594(6)$. By taking into account the two loop
beta function of the model ($\Lambda_L$ is the renormalization group
invariant mass scale corresponding to the bare lattice coupling of the
action~(\ref{actionS})),
\begin{equation}
a(\beta) \Lambda_L = {\rm e}^{-\pi^2\beta}
\left(\frac{1}{2\pi^2\beta}\right)^{5/2}\,
\left[1 + O\left(\frac{1}{\beta}\right)\right]\, ,
\end{equation}
we obtain the ratio $a(\beta_c)/a(\beta=1.5)=0.34(3)$ where the error comes
from the imprecision in the determination of $\beta_c$ (notice that we use
indistinctly the notation $a$ or $a(\beta)$). The error caused
by the ignorance of higher loop terms in the beta function is rather small
within the narrow interval between
$\beta=1.5$ and $\beta_c$ (actually the sole one loop term
is enough to account for about the 85\% of the variability
of the function $a(\beta)$ in that interval).

Since our simulations have been performed at $\beta=1.5$ on a lattice
of time size equal to~6, the temperature of our system was
$T/T_c= a(\beta_c)/a(\beta=1.5) = 0.34(3)$
and the resulting value for the critical chemical potential was
$\mu_c/T_c=6\times (T/T_c)\times 0.175(5) =0.357(10)(32)$
where the first error is due to the Monte Carlo determination of the
critical value of $a\mu$ and the second one derives
from the error in~$\beta_c$.

{}From Table~1 we can
also give an approximation for the critical density. Since the transition lies
between $a\mu=0.17$ and $a\mu=0.18$, from Table~1 we can approximate the
density at the transition point as $a^3\rho_{B,c}=0.0055(15)$ which is the
average of $a^3\rho_B$ among the two values of $a\mu$.
{}From this average we obtain 
$\rho_{B,c}/T_c^3=6^3\times \left(T/T_c\right)^3\times 0.0055(15)=0.047(13)(12)$
where the first error comes from
the lattice determination of $a^3\rho_{B,c}$ and the second one
is due to the determination of~$\beta_c$.

Based on the large number of fermions in our lagrangian and on the fact
that they are quite heavy, we surmise that the critical temperature for
our system at zero chemical potential is $T_c=100-200$~MeV. Then the
value of the lattice spacing at $\beta_c$ is
$a(\beta_c)=1/6T_c=0.00111(^{+56}_{-28})$~MeV$^{-1}$. From the ratio
$T/T_c$ we obtain the value of the lattice spacing in our simulation runs,
$a(\beta=1.5)=0.00327(22)(^{+165}_{-83})$~MeV$^{-1}=0.64(4)(^{+33}_{-16})$~fm
where the first error is due to the imprecision on $\beta_c$ and the
second one to the estimated inaccuracy on $T_c$.

We see that our lattice is coarse. It is a consequence of the small
lattice size (due to computer limitations) and of the need
to work at $T<T_c$ in order to run into a phase transition while $\mu$
is increasing.

Notice that within the above limits for $T_c$, the value of the
critical fermion chemical potential $\mu_c$
ranges from 36(1)(3) to 72(2)(6) MeV. 
Then the critical baryonic chemical potential $\mu_{B,c}=2\mu_c$
varies from 70 to 145 MeV at a temperature $T/T_c =  0.34(3)$,
indicating a curvature of the phase boundary
in the $\mu$--$T$ plane quite larger than what is measured in QCD
with three colours: that does not come as a surprise,
since in two colour QCD
baryon states are degenerate with mesons~\cite{pauligursey},
so that the critical chemical potential at zero temperature
may well be lower than in usual QCD.



\vskip 2cm

\centerline{{\bf Table 2}}
\centerline{Measurements of 
$\langle\overline{\psi}\psi\rangle_{\rm full}$ and $P$ for
varying $\beta$ at $\mu=0$.}
\vskip 2mm
{\centerline{
\vbox{\offinterlineskip
\halign{\strut
\vrule \hfil\, $#$ \hfil \, &
\vrule \hfil\, $#$ \hfil \, &
\vrule \hfil\, $#$ \hfil \, \vrule \cr
\noalign{\hrule}
\beta &
\langle \overline{\psi}\psi\rangle_{\rm full} &
P \cr
\noalign{\hrule}
1.525 &
0.5521(19) &
0.030(1) \cr
\noalign{\hrule}
1.55 &
0.5080(25) &
0.038(1)\cr
\noalign{\hrule}
1.5625 &
0.4828(25) &
0.0402(35) \cr
\noalign{\hrule}
1.575 &
0.4466(26) &
0.0549(23) \cr
\noalign{\hrule}
1.5875 &
0.4131(30) &
0.0685(10) \cr
\noalign{\hrule}
1.6 &
0.3735(22) &
0.0880(18)\cr
\noalign{\hrule}
1.6125 &
0.3454(24) &
0.1041(20)\cr
\noalign{\hrule}
1.625 &
0.3279(15) &
0.1082(20)\cr
\noalign{\hrule}
1.65 &
0.3005(8) &
0.1208(10)\cr
\noalign{\hrule}
1.675 &
0.2801(7) &
0.1315(7) \cr
\noalign{\hrule}
}}
}}

\vskip 1cm


By allowing the valuation of $T_c$ to vary in such a wide range
(100--200 MeV) we have mimicked an estimate of the systematic
error derived from the great number of quark species and their big mass.
There are however other sources of systematic errors which have not been
contemplated in the present analysis: the (wrong) gauge group, the large
value of the lattice spacing, the modest size of the lattice volume and
the use of a non--exact updating algorithm.
The experience in simulating and comparing the physics of SU(2) and SU(3)
gauge theories teaches us that the first problem is possibly
unimportant with regard to the behaviour of the topological susceptibility.
As described above, the main influence of the wrong gauge group 
regards the numerical values of the physical parameters.
The second~and

\begin{figure}
\centerline{\includegraphics[width=\columnwidth]{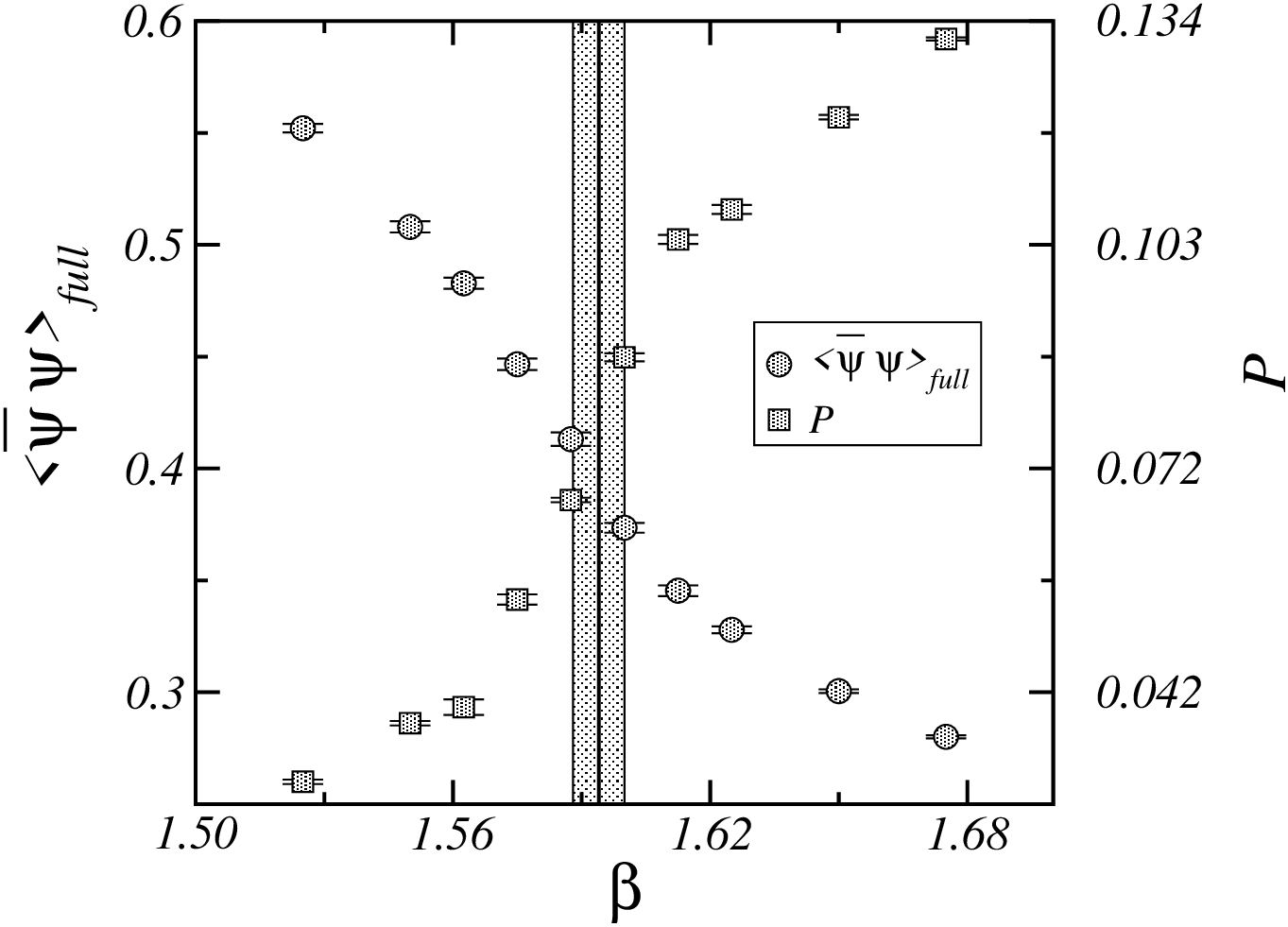}}
\vspace*{12pt}
\caption{Chiral condensate (circles and left vertical axis)
and Polyakov loop (squares and right vertical axis) as a function of
$\beta$ at vanishing chemical potential in the neighbourhood of
the finite temperature transition. The vertical line and band are
the value of $\beta_c$ and its error.}
\end{figure}

\noindent third problems can be ameliorated by studying the theory
on larger volumes.
We think that none of these errors can change the two main
conclusions of our paper: the drop of the topological
susceptibility at the transition and the coincidence of all transitions. 

\subsection{Relationship between transitions}

We have claimed that the three transitions studied in the present
paper, namely the one related to topology, to the
Polyakov loop and to the chiral condensate are
concomitant: they all happen at about the same value of the
chemical potential. We substantiate this statement with the
help of the plot shown in Fig.~6. In this plot we display the
derivatives

\begin{figure}
\centerline{\includegraphics[width=\columnwidth]{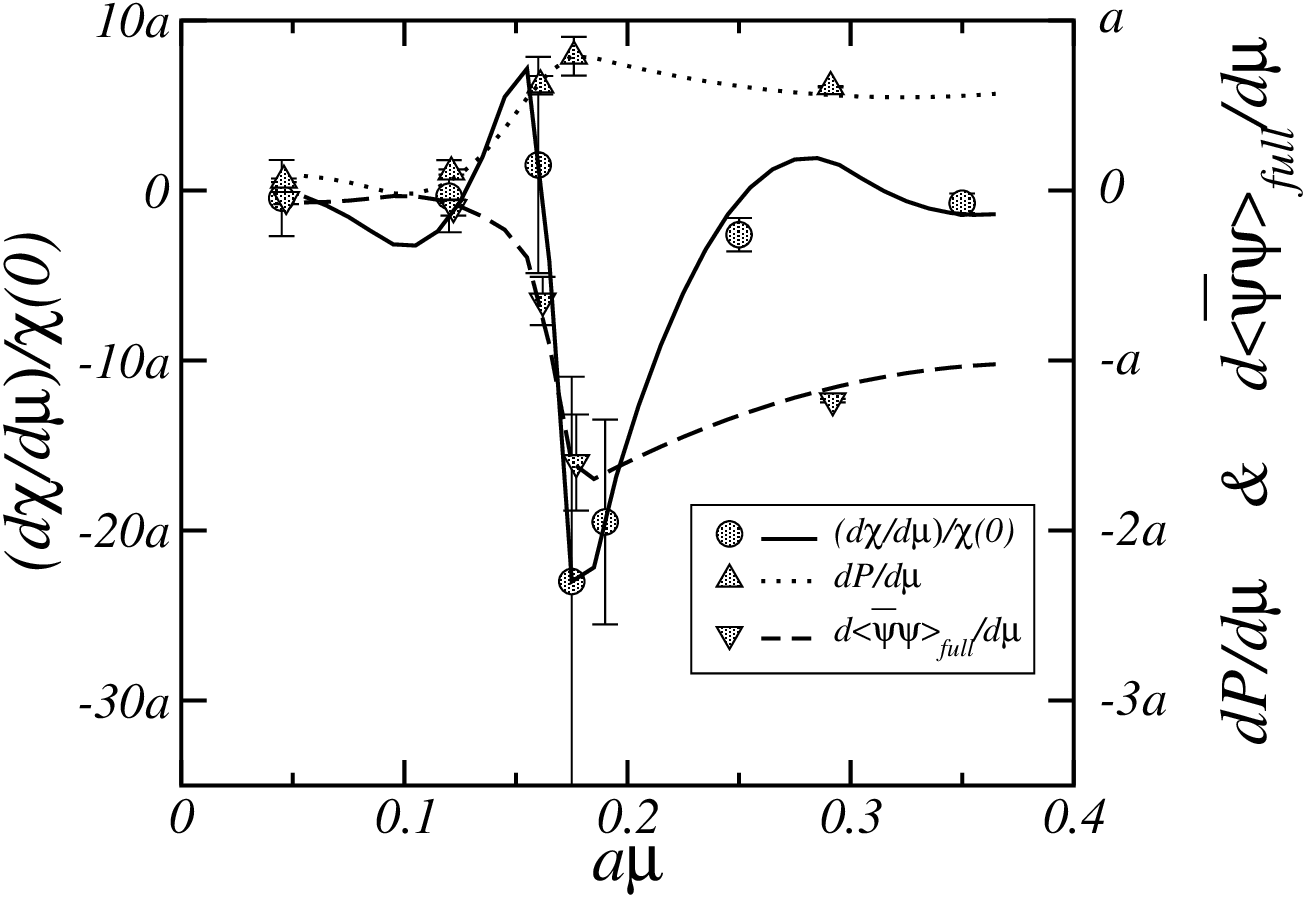}}
\vspace*{12pt}
\caption{Derivatives with respect to the chemical
potential $\mu$ of the normalized topological susceptibility
(circles, continuous line and left vertical axis), Polyakov loop (up triangles,
dotted line and right vertical axis) and chiral condensate (down triangles,
dashed line and right vertical axis). The two vertical axes are expressed in units of
the lattice spacing $a$. The three sets of data have been slightly
shifted horizontally to avoid the overlapping of various symbols and
error bars. Lines are the result of the interpolation described
in the text.}
\end{figure}
\noindent of the three observables (with immaterial proportionality
constants) with respect to the chemical potential. If the transitions
are identified as the points where the variations are sharper then
the Figure seems to give unquestionably evidence in favour of the coincidence
for the three transitions. This conclusion is the second main result
of the paper.

The plot is shown within the interval of 
chemical potentials spanning from 0 to the maximum
$\mu$ where no Pauli blocking effects are apparent.
The isolated points (circles, up and down triangles) for the observable
$O$ ($\chi(\mu)/\chi(0)$, $P$ or
$\langle\overline{\psi}\psi\rangle_{\rm full}$)
have been obtained by calculating the discrete derivative
\begin{equation}
\frac{O(\mu_i) - O(\mu_{i-1})}{\mu_i - \mu_{i-1}}
\end{equation}
from the data in Table~1 and placing the result at position
$a(\mu_i + \mu_{i-1})/2$. Furthermore Monte Carlo
data for each observable have been interpolated
by a natural cubic spline~\cite{numrecipes}. Then the derivative of these
interpolation functions have been calculated and they are the lines plotted in Fig.~6.
All three lines and all three sets of data points show a clear peak at
the same value $a\mu_c=0.175$. The continuous
line, which corresponds to the topological susceptibility,
displays other points with a large derivative. This is an
spurious effect on the spline interpolation due to the fluctuation that
makes the value of $\chi_L$ at $a\mu=0.17$ to move upward a little bit,
see Fig.~1.

A similar figure can be drawn for pure gauge theories and for full QCD indicating
that the concomitancy is a common feature of the three kinds of transition
for all theories that we have studied up to date~\cite{alldelia}.

\section{Conclusions and outlook}

\vskip 5mm

We have studied on the lattice the behaviour of the topological
susceptibility $\chi$ in QCD with two colours and 8 flavours of
quarks as a function of the quark chemical potential
$\mu$ at finite temperature. The so--called sign problem
hinders a similar investigation in full QCD with three colours.

Previous studies of topology in Yang--Mills theories at finite
temperature for two and three colours both with or without fermions
have shown that $\chi$ undergoes an abrupt drop at the deconfinement phase
transition. The results of our
paper indicate that a similar quick drop takes place also at the finite
density phase transition, (see inset in Fig.~1). It happens at a
value of the chemical potential $a\mu_c = 0.175(5)$.
This value of $\mu$ is the same at which the Polyakov loop and the
chiral condensate undergo sudden changes (see Fig.~2 and Fig.~4).
The rise of the Polyakov loop demonstrates enhanced screening
possibly compatible with deconfinement and the chiral condensate
exhibits a behaviour that is compatible with
a restoration of flavour chiral symmetry (see Fig.~4).

If the chemical potential is further increased, the topological susceptibility rises
again as more and more fermions are inserted into the system.
We have given compelling arguments showing that this rise is not
the signal of a new physical phase but just an artifact of the
UV regularization of the lattice, which makes the number of
accessible fermion levels finite. Indeed at the saturation point
$\mu_s$ (for our choice of parameters $a\mu_s\approx 1.2$)
the system is completely packed with fermions whose dynamics,
due to the Pauli principle, gets frozen. Therefore for $\mu>\mu_s$
the gluons are the only dynamical degrees of freedom and the
system behaves as the pure gauge theory at the same value of the gauge
coupling and consequently the topological susceptibility takes the
corresponding quenched value. In order
to support these assertions, we have calculated the baryonic density and verified that it
becomes equal to 1 baryon per site for all $\mu > \mu_s\,$, (see Fig.~3).
Similarly, also other
observables like the Polyakov loop (Fig.~2), the average plaquette (Fig.~3) and the
chiral condensate (Fig.~4) take their quenched values for $\mu > \mu_s\,$.

We stress that we defined $\mu_s$ as the value of the quark chemical
potential where saturation becomes complete or, stated otherwise,
where the theory becomes quenched.

Since the Pauli saturation of fermions is a pure lattice effect with
no counterpart in the continuum, we have not made a detailed study to
establish whether its onset corresponds to a true (unphysical) phase
transition or not: that may be the subject of further investigation.
In fact $\mu_s$ must diverge in the continuum limit. Unfortunately
this lattice artifact precludes us from making contact with
the physics of extremely high densities by numerical simulations.

We have shown that a drop in the signal of the topological susceptibility
occurs in the theory with
two colours and 8 equal mass flavours. What about the true theory? Due to the
similarities in the topology of SU(2) and SU(3) it is conceivable to expect
that the drop must be a genuine fact of QCD. Instantons
are responsible for the topological properties and they are present in
all SU($N$) gauge models.

We call $T_c$ the temperature where the chiral vector symmetry is restored
at vanishing chemical potential. By resorting to the above--mentioned
similarities between the present model and the true theory of the strong
interactions, we have assumed that $T_c$ ranges between 100 and 200~MeV.
{}From this approximation and knowing the value $\beta_c$ of the gauge coupling
at which the chiral symmetry is restored for $\mu=0$ (see Fig.~5),
we have extracted an estimate of the lattice spacing in physical units. Also
the ratio $\mu_c/T_c=0.357(10)(32)$ has been calculated,
where the first error is due to the lattice determination of
$a\mu$ at the critical point and the second one 
is derived from the error band in $\beta_c$ (see Fig.~5).
Furthermore $T/T_c= 0.34(3)$. Notice that this ratio is insensitive
to the imprecision on $T_c$. Moreover the baryon density $\rho_B$ at
the critical point is
$\rho_{B,c}/T_c^3 = 0.047(13)(12)$ where 
the errors have analogous meanings than those for $\mu_c/T_c$:
lattice evaluation of $\rho_{B,c}$ and error on $\beta_c$ respectively.
All the above numerical results are subject to several sources of
systematic errors (wrong gauge group, small lattice size, large
lattice spacing). Hence we
consider them as rough estimates of the true values. However none
of the preceding sources of systematic errors is expected to be able
to influence the main conclusions of the present paper, namely
the drop in the topological susceptibility and the coincidence of this
drop with the chiral and Polyakov loop transitions.

By using the assumed limits to the value of the critical temperature
at zero chemical potential $T_c$, the critical baryonic chemical
potential $\mu_{B,c}=2\mu_c$ can be estimated to range from 70 to
145 MeV at $T/T_c=0.34(3)$. Therefore the value of $\mu_{B,c}$
at zero temperature may be quite lower than what is expected in QCD
with three colours. That is not a surprise since in the theory with two
colours the lightest baryon state is degenerate with the pion.

We have not studied the order of the transition and the properties
of the states at both sides of this transition.
On the other hand the number and mass of the quarks
in our study are unphysically large.
These are important aspects which deserve further analysis because
the order of the transition might depend on these parameters.

Our results indicate that the screening of colour forces, 
chiral symmetry and topology are interrelated
in the $T\approx (0.3-0.4) T_c$ region of the two colour QCD phase diagram~\cite{Kogut02}. 
It would be interesting
to repeat this study at other temperatures, in particular at very
small ones where the transition goes from the hadronic
to  the superfluid phase,
which is still confining~\cite{Toublan5,Skullerud5},
and where chiral symmetry remains broken by a diquark condensate.
Such low temperatures, say $T\ltapprox 0.1\,T_c$, were not accessible
to our simulations because on our lattice sizes it would have required
such a large value of the lattice spacing that the results would
probably have lost all physical meaning.

\section{Acknowledgements}

\vskip 5mm

It is a pleasure to thank Adriano Di Giacomo for useful
comments and discussions. B.A. and M.P.L. also thank Massimo Testa
for discussions and B.A. ackowledges Giancarlo Cella for suggestions about
the statistical treatment of data.
We also thank Michele Pepe for a critical reading of
the original draft of the paper and for invaluable help in the
simulation runs.



\end{document}